\documentclass{ws-p9-75x6-50}

\def\seteps#1#2#3#4{\vskip#3\relax\noindent\hskip#1\relax
 \special{eps:#4 x=#2, y=#3}}
\def\centereps#1#2#3{\vskip#2\relax\centerline{\hbox to#1{\special
  {eps:#3 x=#1, y=#2}\hfil}}}

\def\s|{\left|\!\left|}
\def\d|{\right|\!\right|}
\def\Hcal{{\cal H}}

\def\bfHcal{{\mbox{\boldmath$\Hcal$}}}

\def\be{\begin{equation}}
\def\ee{\end{equation}}
\def\ie{{\it i.e.},\ }
\def\eg{{\tt e.g.},\ }
\begin{document}

\title{
Fast instability indicator in few dimensional dynamical systems.
}

\author{Piero Cipriani}

\address{\it Istituto Nazionale di Ottica Applicata, L.go E.Fermi, 6  - 50125 - Firenze (Italia)\\ 
{\rm and}\\
 {\sf I.C.R.A.} - Coordinating Center - P.le della Repubblica, 10 - 65121 Pescara (Italia)\thanks{
Permanent {\it postal} address: {\sf C.S.S.} - via E.Nardi, 14-16 --
02047 -- Poggio Mirteto (RI) Italia ; E-mail: cipriani@icra.it
}\\
E-mail: cipriani@ino.it
}

\author{Maria Di Bari}

\address{\it Dip. di Fisica, Univ. di Parma - Parco Area delle Scienze, 7/A - 43100 Parma (Italia)\\ 
{\rm and}\\
{\sf I.N.F.M.} - UdR di Parma\\
E-mail: mariateresa.dibari@fis.unipr.it
}

\maketitle
\abstracts{Using the tools of Differential Geometry, we define a new {\it fast} chaoticity indicator,
able to detect dynamical instability of trajectories much more effectively, (\ie {\it quickly}) than the usual tools,
like Lyapunov Characteristic Numbers (LCN's) or Poincar\'e Surface of Section.
Moreover, at variance with other {\sl fast} indicators proposed in the Literature, it
gives informations about the asymptotic behaviour of trajectories, though being
{\it local} in phase-space. Furthermore, it detects the chaotic or regular nature of 
geodesics without any
reference to a given perturbation and it allows also to discriminate
between different regimes (and possibly sources) of chaos in distinct
regions of phase-space.
}
Chaotic dynamics is believed to be the rule rather than the exception in any generic
nonlinear dynamical system; however, there is no general {\it fast} method able to determine
what's the fraction of the allowed phase space occupied by chaotic orbits.
This problem exists in the case of few degrees of freedom (dof) systems
and for many dimensional ones as well, though it originates from different causes
in the two situations. 
Recently the authors derived\cite{HHPRL} a {\sl geometric indicator} of Chaos, able to
single out the chaotic or regular character of individual orbits simply computing a pair
of {\sl short time correlation functions} of geometric quantities defined on the tangent bundle
of the system.
Here we discuss some points which, in our opinion, make our indicator preferable with respect 
to other {\sl "fast"} indicators, and, from a practical viewpoint, we compare its {\sl performances}
with those of the others, showing the faster convergence properties and the richer
amount of information it contains.

In addition to the well known reduction of dynamics to geometry which can be accomplished
on the grounds of the Maupertuis principle of Least Action\cite{Arnold}, there is a variety
of possibilities of transcription of dynamical trajectories into geodesic flows over suitable
manifolds.
A particularly appealing choice, in the case of generic Lagrangian systems, with 
a more general than simply quadratic dependence
on velocities, is given by the Finsler approach, which
furnishes a metric on the tangent bundle of the Lagrangian system under examination\cite{PSS2,PRE97}.
Dynamical instability is linked to curvature properties of the manifold
through {\sl Jacobi--Levi-Civita} equation for geodesic spread (or a suitable generalization thereof), which 
describes the evolution of a generic perturbation $\delta q^a$ to a given geodesic, as a function
of the affine parameter, $s$, along the geodesic itself:
\be
{{\nabla} \over {ds}}\left({{\nabla \delta q^a} \over {ds}}\right)
 + {\cal H}^a{}_c \delta q^c = 0, 
\label{EDG}
\ee
where the {\sl stability  tensor}\cite{CPTD,PSS1}, ${\cal H}^a{}_c$, defined along any geodesic,
is related to the Finsler curvature tensor, $K^a{}_{bcd}$, associated to the metric, by
\be
{\cal H}^a{}_c \doteq  K^a{}_{bcd} x^{\prime b} x^{\prime d}\ ,
\ee
and $x^\prime$ is the unit tangent vector to the geodesic. Finsler curvature tensor (and consequently 
the stability tensor) is a pretty complicated object 
\cite{PSS2,PRE97}; however the main features of the geodesic flow depend essentially
on suitable {\sl averages} and moments of scalars built from ${\cal H}^a{}_c$.
In the case of $N$-dof systems, the curvature properties of the associated Finsler manifold 
are described by the $N$ non trivial {\sl principal sectional curvatures}\cite{PSS1}, 
$\{\lambda_i (s)\}$, which {\sl are specific of each given path}, \ie  depend not only on the 
point on the configuration space but also
on the tangent vector along the given geodesic\footnote{This is one of the main differences
which make the Finsler geometrization more adeherent to the complete dynamics than others, like
Jacobi or Eisenhart.}.
If we restrict to two dof systems, the curvature properties of the Finsler manifold are then described
by the two principal sectional curvatures $\{\lambda_1,\lambda_2\}$. 
The {\sl mean} curvature
$\kappa$ and the anisotropy $\vartheta$ are defined, in terms of the principal
sectional curvatures, simply as
\be
\kappa\doteq {{\lambda_1+\lambda_2}\over {2}} \equiv {{{\rm Tr}(\bfHcal)}\over {2}}
= {{{\sf Ric}_{\rm F}(x^{\prime})}\over {2}}
\ \ \ ;\ \ \  
\vartheta\doteq {{\lambda_1-\lambda_2}\over {2}}\ .
\ee
where it has been exploited the link between the (Finsler generalization of)
Ricci curvature along the flow and the trace of the stability tensor.
For any realistic dynamical system, it is easy to verify that the 
associated Finsler manifold and the geodesics on it are characterized by 
principal sectionals which fluctuate around a well defined
average\footnote{The only requirement is that the phase space is bounded and
the dynamical system is singularity-free; though with some cautions
the application of our approach can be extended even to singular lagrangians.}, 
$\lambda_i=\lambda_i(s)$.
Schur theorem\cite{SyngeSchild} establishes a link
between the anisotropy of a manifold and the fluctuations of the curvatures moving along a given
path, in particular along a geodesic, and suggests a possible tool to characterize
the interplay between them.

Introducing the correlation function of any two given observables,
$X(s)\doteq \tilde{X}[q(s),q^\prime(s)]$ and $Y(s)\doteq \tilde{Y}[q(s),q^\prime(s)]$
computed  {\sl over a finite {\it interval}}, $S$, along a geodesic 
with initial conditions
$({\bf q}_{{}_0},{\bf q^\prime}_{{}_0})\equiv [{\bf q}(0),{\bf q^\prime}(0)]$,
\be
\widetilde {\cal C}_S [X,Y] ~\doteq~ 
{{\langle X\cdot Y\rangle_S}\over{\left(\langle X^2\rangle_S~\cdot~\langle 
Y^2\rangle_S\right)^{1/2}} }\ ,
\ee
we define the functional $R_F[S]$ over the given segment of the geodesic,  as
\be
R_F[S]~\equiv~R_F({\bf q}_{{}_0},{\bf q}^{\prime}_{{}_0})~\doteq~{
{\widetilde {\cal C}[\vartheta, \delta\vartheta]}
\over
{\widetilde {\cal C}[\kappa, \delta\kappa]}
}~\geq~0\ ,
\ee
where $\delta X(s)~\doteq~ X(s) - \langle X\rangle$.
We have shown\cite{HHPRL,HHPRE} that the
stability properties of the geodesic flow, and then the degree of chaoticity of
the trajectories, possess an one-to-one correspondence with the value assumed
by this functional over any short segment of a geodesic. Just by computing
$R_F$ over a few periods along an orbit, suffices to have a reliable indication on its
regular or chaotic character.
Here we report some supplementary results, still referred to the {\sl paradigmatic}
H\'enon--Heiles system\cite{HH}, whose Lagrangian is
\be
L({\bf q},{\bf \dot{q}}) = {1\over{2}} \left(\dot{q}_x^2 + \dot{q}_y^2\right)
 -   \left[ \frac{1}{2} \left( x^2 + y^2\right) + x^2y - {{y^3}\over{3}}\right]\ ,
\ee
which emphasize the much faster convergence of our indicator with respect to the usual
ones.
A pictorial taste of the reliability and effectiveness of the Finsler indicator, 
has been given, for example, in figures 1 and 2 of ref.\cite{HHPRL}, where
we made a synoptic comparison of the gray-scale map of $R_F$ with the PSS 
for a typical energy value $(E=0.125)$, for which the phase space of this system
is almost equally populated by chaotic and regular trajectories; 
noticing that the integration time $S$, used to produce
the map of $R_F[S]$-map is there about two orders of magnitude shorter than the one required
to obtain the PSS with a comparable resolution level. 
Furthermore, we observe
that, if a better resolution of the substructures on the surface is sought, the computation
of $R_F$ requires obviously only a more careful numerical integration, whereas
the PSS needs also to improve the bisection algorithm needed to find the
exact intersection point.
More detailed results and discussions, and extensive applications to other few dimensional systems 
can be found elsewhere\cite{HHPRE}. 
\begin{figure}
\seteps{-.7cm}{7cm}{11cm}{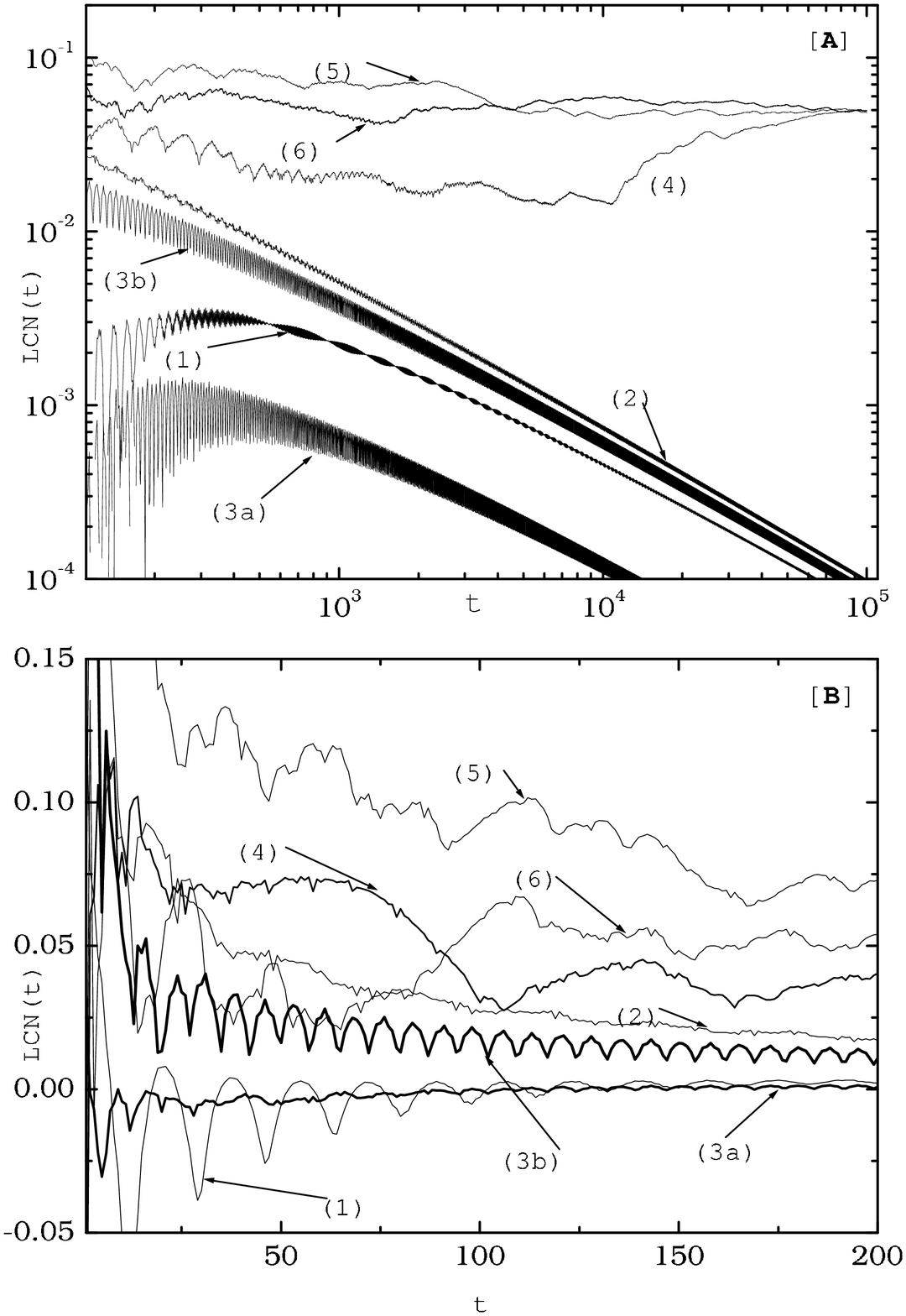}
\vskip-11.9cm
\seteps{6.1cm}{7cm}{11.9cm}{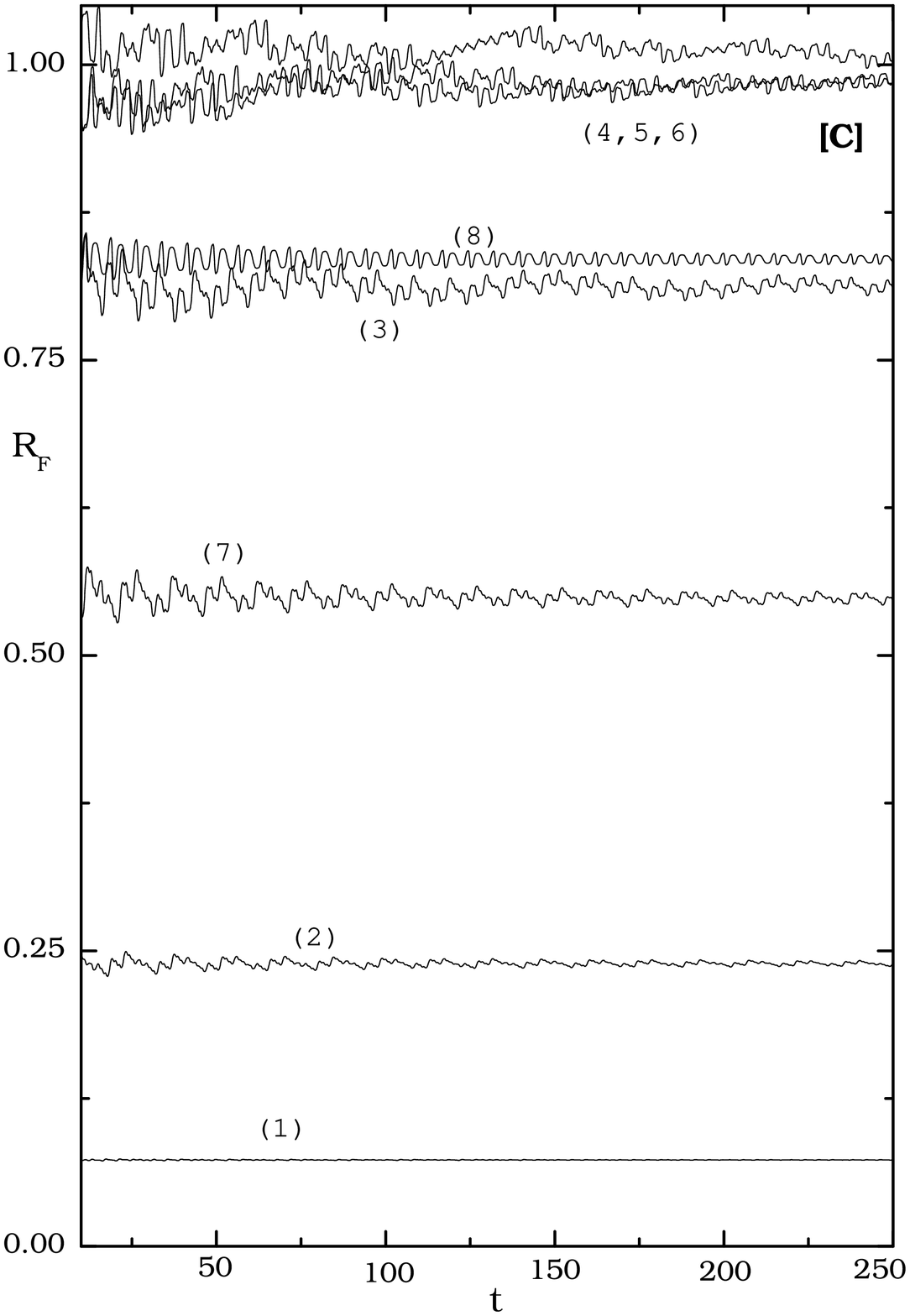}
\vskip-.8cm
\caption{Convergence properties of LCN's and of the (normalized) geometric indicator $R_F$: 
panels $[A]$ and
$[B]$ display respectively the long and short time behaviours
of LCN's for a set of initial conditions on the energy surface $E=0.125$. Panel $[C]$ shows
the values of $R_F$ for the same trajectories and for two further regular orbits.
The much faster convergence of the latter indicator is evident.}
\end{figure}
To further evaluate quantitatively the fast convergence of the proposed indicator, 
we report in figure 1 the values of the LCN's and of $R_F[S]$ as a function of {\it time}\footnote{
See \cite{HHPRL,PSS2} for the relationship between the Newtonian time $t$ and the Finsler
affine parameter $s_F$. We recall only that we can always choose, as we did here, $s_F(t)\cong t$.}
for some typical orbits of the same system, which makes self-evident the much faster signature of
regular or chaotic motion obtained using the geometric indicator.
In addition, we stress also that the values of our indicator are completely 
independent from any given choice of the initial perturbation vector, whereas this is not true
for the computation of the LCN's, whose values, especially for few dof systems, can be strongly
influenced by the choice of the orientation of the initial disturbance vector. This
is mostly evident if the reference orbit is near the border between regular islands
and the stochastic sea, where {\it stickness} phenomena are expected to occur.
The three plots in the figure indeed demonstrate all what has been said here:
\begin{itemize}
\item Plot $[A]$, showing the long time behaviour of LCN's, singles out that 
for some rather {\sl sticky} orbits, the
time required for the non vanishing LCN's to converge towards the {\it true}
asymptotic values can be about three orders of magnitude larger than the
associated Lyapunov time, $\tau_L\sim LCN^{-1}$. 
\item Moreover, the two curves labeled as $(3)$ and $(3b)$ show that up to
$t\cong 10^3$ the behaviour of two LCN's, computed along the {\sl same orbit} but with
a different choice of the orientation of the perturbation vector in phase space, can display
completely different behaviours.
\item This is confirmed by the behaviour reported in the plot [B], where the same curves of the 
previous graph are restricted to a relatively short interval, which reveals how, up to $t\sim 200$,
LCN's are completely inadequate to discriminate between regular and chaotic orbits.
\item Plots in [C] show instead that the (normalized\cite{HHPRL}) geometric indicator $R_F$ possess
much better convergence properties: it is evident that already at $t\sim 50$ the discrimination
between chaotic and regular orbits is clear and the fluctuations around the average for 
regular orbits are almost negligible. Moreover the values assumed by our indicator give
a measure about the {\sl degree of regularity} of non chaotic trajectories, indicating which
{\sl topological region} of phase space will become chaotic {\it earlier}, as we increase
the perturbation parameter, \ie in this case, the energy.
\end{itemize}
A couple of final comments are in order. 

The first one addresses the {\sl dichotomy} encountered
in the characterization of dynamical systems, where no intermediate status exist between
{\sl ordered} and {\sl chaotic} motions, for a given dynamical system: the picture
based on the usual tools shows that the change from asymptotically ordered to chaotic dynamics,
presents itself indeed as a {\it phase transition}, along with a variation of an external
parameter (\eg the energy), with a sudden jump of asymptotic {\sl observables}
(\eg LCN's) from a vanishing value to a finite one.
The proposed geometric indicator, though confirming the occurrence of a {\sl change of
state} phenomenology, brought to evidence even by local correlation functions, indicates however 
the presence of gradual, premonitory changes of local observables, accompanying the
approach to the {\sl critical point}.

The second remark deals with some recently proposed so-called {\sl fast indicators}, like
{\sl rotation angles} and {\sl Fast Lyapunov Indicators}\cite{Fro,Con}. 
These interesting alternative tools to detect
chaoticity, when applied to continuous time dynamical systems, give {\sl "fast"}
informations about the {\sl "local"} dynamics; if, inseatd, an asymptotic characterization
is sought, then their computation needs, in general,
as much as integration time as that required by the evaluation of LCN's.
In addition they also make necessarily reference to a given choice
of the perturbation vector. Nevertheless they are best suited for the
study of the KAM and Nekhoroshev thresholds in the case of discrete time maps, where our
approach is instead unsuitable.

In any case, we point out that our {\sl geometric} indicator, which is defined only in terms 
of the region of the tangent bundle explored by a trajectory, gives a deeper
conceptual insight on the sources of Chaos in dynamical systems, just because
is strictly linked to the properties of the underlying manifold and make
no reference to any perturbations.

\end{document}